\documentclass[a4paper,
              ]{article}
\usepackage[USenglish]{babel}

\usepackage{subcaption}
\usepackage{graphicx}
\usepackage{siunitx}
\usepackage{amsmath}
\usepackage{amsfonts}
\usepackage{amssymb}

\begin{document}
\title{Flat Bunches with a Hollow Distribution for Space Charge Mitigation}

\author{A. Oeftiger\thanks{adrian.oeftiger@cern.ch, also at \'Ecole Polytechnique F\'ed\'erale de Lausanne, Lausanne, Switzerland},
        H. Bartosik, A. Findlay,
        S. Hancock, G. Rumolo \\
        CERN, Meyrin, Switzerland
        }

\maketitle

\begin{abstract}
   Longitudinally hollow bunches provide one means to mitigate the impact of
   transverse space charge. The hollow distributions are created via dipolar
   parametric excitation during acceleration in CERN's Proton Synchrotron
   Booster. We present simulation work and beam measurements. Particular
   emphasis is given to the alleviation of space charge effects on the
   long injection plateau of the downstream Proton Synchrotron machine,
   which is the main goal of this study.
\end{abstract}

\section{Introduction}
In order to push brightness limits given by transverse space charge effects,
one can modify the longitudinal bunch shape. Reducing the peak line
charge density decreases the space charge tune spread and the beam becomes
less prone to betatron resonances located near the working point.
The standard approach to flatten the bunch profile makes use of double
harmonic RF systems in bunch lengthening mode.
In this paper we show that hollow longitudinal distributions provide a viable
alternative.

Circular accelerators usually feature the strongest space charge impact at injection.
Ideally, bunches should arrive with an already flattened longitudinal profile.
Hollow bunches can be created in the upstream accelerator and then
transferred into a single harmonic RF system.

Our experiment applies this concept to the CERN Proton
Synchrotron Booster (PSB) which provides beams to the Proton Synchrotron (PS).
For the double-batch filling scheme, first four bunches are injected into the PS
and circulate at 
the injection energy while
the second batch is being prepared in the PSB. After \SI{1.2}{\second} the second
batch is injected and the PS acceleration ramp starts. During this period,
transverse space charge effects can result in transverse emittance growth and/or
beam losses and therefore become a performance limitation for high brightness
LHC beams \cite{wasef}.

We present a reliable procedure to create hollow distributions during the PSB
acceleration ramp involving minimal changes to the current operational cycle.
Finally, we compare emittance blow-up during the PS injection plateau
between these hollow bunches and standard parabolic bunches.
These efforts build on the experience from past hollow bunch experiments
\cite{cappi1993measurement,garobyhancock}.

\section{Theoretical Considerations}
For a transversely Gaussian normal distributed bunch of particles, the
detuning effect of the beam self-fields can be quantified in terms of the
transverse space charge tune spread
\cite{schindl},
\begin{equation}
    \Delta Q_u(z) = -\frac{r_p \lambda(z)}{2\pi\beta^2\gamma^3}\,
    \oint\,ds \frac{\beta_u(s)}{\sigma_u(s)\,\left(\sigma_x(s)
                                    + \sigma_y(s)\right)}
    \label{eq: gauss tune spread}
\end{equation}
with $u=x$ or $u=y$ for the horizontal resp.\ vertical plane, $z$ denoting
the longitudinal position with respect to the beam centre-of-gravity,
$\lambda(z)$ the line charge density in \si{\coulomb/\meter},
$r_p$ the classic particle radius, $\beta$ the speed in units of speed of light,
$\gamma$ the Lorentz factor,
$\beta_u(s)$ the betatron function depending on the longitudinal location
$s$ around the accelerator ring and $\sigma_u(s)$ the corresponding
transverse beam sizes. In presence of dispersion $D_x(s)$,
the momentum distribution contributes to the horizontal beam size.
Assuming also the momentum distribution to be
Gaussian normal distributed yields the well-known expression
\begin{equation}
    \sigma_x(s) = \sqrt{\frac{\beta_x(s)\epsilon_{x}}{\beta\gamma} +
                        D_x^2(s) \delta_\text{rms}^2} \quad ,
    \label{eq: gauss beam size}
\end{equation}
where $\epsilon_x$ is the normalised beam emittance and $\delta_\text{rms}$ the
root mean square of the relative momentum distribution. NB:
Eq.\ \eqref{eq: gauss beam size} is no longer valid for beams with
a momentum distribution that significantly deviates from a Gaussian.

Longitudinally hollow phase space distributions address two aspects of Eq.\
\eqref{eq: gauss tune spread} to reduce $\Delta Q_u^\text{max}$ compared
to Gaussian or parabolic bunches. They project to
\begin{enumerate}
    \item intrinsically flat bunch profiles (reduced $\lambda_\text{max}$) and
    \item broader momentum profiles (increased $\delta_\text{rms}$ and $\sigma_x$).
\end{enumerate}

To create such hollow bunches during the PSB cycle, a longitudinal dipolar
parametric resonance is excited by phase modulation \cite{lee}. To this end we use the
phase loop feedback system, which aligns the RF reference phase
$\phi_\text{rf}$ with the centre-of-gravity of the bunch. We modulate the phase
loop offset around the synchronous phase $\phi_S$:
\begin{equation}
    \phi_\text{rf}(t) = \phi_S + \hat{\phi}_\text{drive}\sin(\omega_\text{drive} t) \quad .
\end{equation}
To excite the beam, the driving frequency $\omega_\text{drive}$ needs to satisfy
the resonance condition
\begin{equation}
    m\, \omega_\text{drive} \simeq n\, \omega_S \quad ,
\end{equation}
where $\omega_S$ denotes the angular synchrotron frequency.
The integer numbers $m$ and $n$ characterise the $m:n$ parametric resonance.
The actual synchrotron frequency of any particle within the RF bucket decreases
with its synchrotron amplitude due to increasingly non-linear synchrotron motion
towards the separatrix. Below transition, as in the PSB, longitudinal space
charge additionally reduces $\omega_S$.

\section{PSB Hollow Bunch Creation}
\subsection{PyHEADTAIL Simulations}

\begin{figure}[htb] \centering
    \begin{minipage}{\textwidth}
    \begin{subfigure}[t]{0.32\textwidth} \centering
        \includegraphics[width=\linewidth]{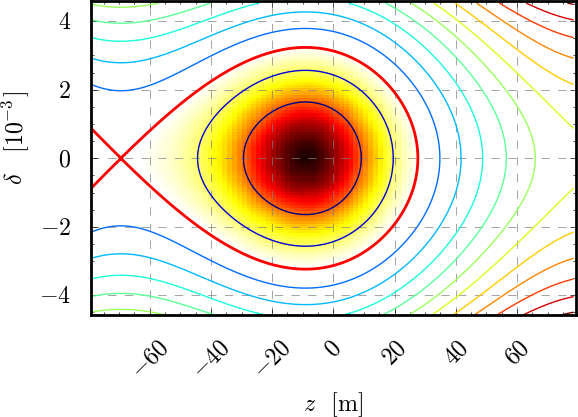}
        \caption{Initial Gaussian bunch}
    \end{subfigure}
    \hfill
    \begin{subfigure}[t]{0.32\textwidth} \centering
        \includegraphics[width=\linewidth]{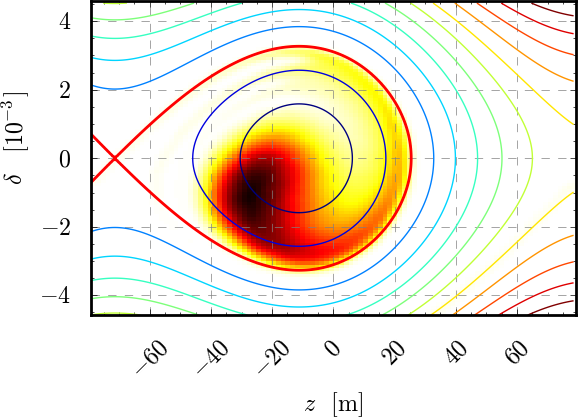}
        \caption{After 2 synchrotron periods}
    \end{subfigure}
    \hfill
    \begin{subfigure}[t]{0.32\textwidth} \centering
        \includegraphics[width=\linewidth]{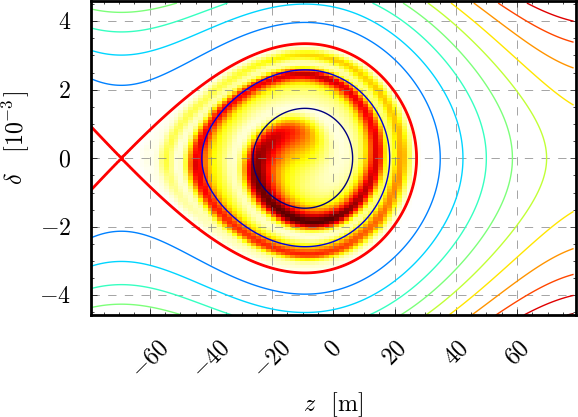}
        \caption{After 6 synchrotron periods}
    \end{subfigure}
    \caption{Longitudinal phase space $(z, \delta)$ with $\delta=(p-p_0)/p_0$
    during the excitation of the 1:1 dipolar parametric resonance in the PSB
    (from PyHEADTAIL \cite{pyheadtail} simulations).}
    \label{fig: simulations}
    \end{minipage}
\end{figure}

By driving the $1:1$ resonance at a frequency slightly below the linear
synchrotron frequency, $\omega_\text{drive}\approx 0.9\omega_{S,\text{lin}}$,
the particles in the bunch core are excited to higher synchrotron amplitudes.
Figure \ref{fig: simulations} shows the depletion of the bunch
centre within a few synchrotron periods leading to hollow longitudinal
phase space distributions. Higher order
resonances create two or more filaments spiralling outward from the bunch
centre and are thus less effective for depletion.

The synchrotron frequency spread between the inner- and outer-most particles
leads to a filamentation-like angular spread. The modulation duration determines the azimuthal
span to which the excited particles surround the depleted bucket centre.
The optimal duration distributing the particles as evenly as possible depends
in descending importance on the excitation amplitude $\hat{\phi}_\text{drive}$,
the ratio between longitudinal emittance and bucket acceptance, and
the beam intensity. The latter dependency becomes evident during intensity scans
and is explained by decoherence suppression due to longitudinal space charge,
which reduces the frequency spread over the particles and may prevent the
filamentation process \cite{thesis}.

The final longitudinal emittance $\epsilon_z$ varies with the bunch intensity,
excitation period and amplitude. To reach a specific $\epsilon_z$,
modifying $\hat{\phi}_\text{drive}$ turns out to be the most effective
parameter, while the excitation duration is fixed beforehand by maximising the
azimuthal phase space distribution.

\subsection{Implementation in PSB}
Based on the current operational LHC-type beam set-up, we introduced the
phase modulation at cycle time C575 (corresponding to an intermediate energy of
$E_\text{kin}=\SI{0.71}{\giga\electronvolt}$) in a single harmonic accelerating
bucket. During \SI{9}{\milli\second} equivalent to 6 synchrotron periods the
beam is driven onto the resonance starting from an initial matched longitudinal
emittance of $\epsilon_{z,100\%}=\SI{1.1}{\electronvolt\second}$. With these
settings, the resulting distributions appeared consistently and reproducibly
depleted.

Varying the driving frequency for the parametric resonance revealed a broad resonance
window. The beam turned out to be correctly excited for frequencies in the range
\begin{equation*}
    \SI{649}{\hertz} \leq \frac{\omega_\text{drive}}{2\pi} \leq \SI{734}{\hertz} \quad .
\end{equation*}
This resonance window is sharply defined up to \SI{1}{\hertz}.

Special attention had to be given to optimise the phase loop gain during the
excitation process: for a too strong gain, the phase loop continuously realigns
the phase of the main C02 cavities with the beam. This counteracts
the excitation and leads to severely perturbed distributions.

\begin{figure}[htb] \centering
    \begin{subfigure}[t]{0.23\textwidth} \centering
        \rotatebox[origin=c]{90}{\small \quad energy $E$}
        \begin{minipage}{0.85\linewidth}\centering
            \includegraphics[width=\linewidth]{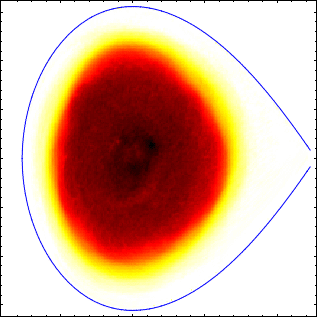} \\
            \small phase $\phi$
        \end{minipage}
        \caption{PSB C573, before excitation}
    \end{subfigure}
    \hfill
    \begin{subfigure}[t]{0.23\textwidth} \centering
        \rotatebox[origin=c]{90}{\small \quad energy $E$}
        \begin{minipage}{0.85\linewidth}\centering
            \includegraphics[width=\linewidth]{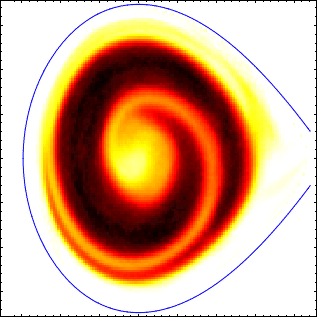} \\
            \small phase $\phi$
        \end{minipage}
        \caption{PSB C591, after excitation}
    \end{subfigure}
    \hfill
    \begin{subfigure}[t]{0.23\textwidth} \centering
        \rotatebox[origin=c]{90}{\small \quad energy $E$}
        \begin{minipage}{0.85\linewidth}\centering
            \includegraphics[width=\linewidth]{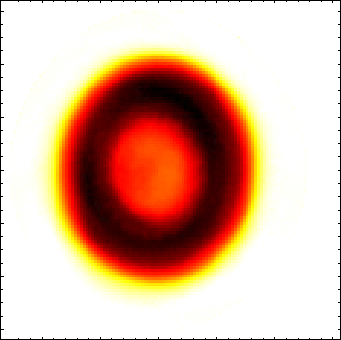} \\
            \small phase $\phi$
        \end{minipage}
        \caption{PSB C800 (\SI{5}{\milli\second} to extract.)}
    \end{subfigure}
    \hfill
    \begin{subfigure}[t]{0.23\textwidth} \centering
        \rotatebox[origin=c]{90}{\small \quad energy $E$}
        \begin{minipage}{0.85\linewidth}\centering
            \includegraphics[width=\linewidth]{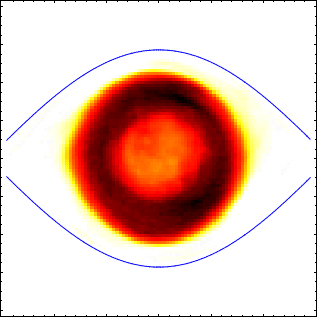} \\
            \small phase $\phi$
        \end{minipage}
        \caption{PS C171 (\SI{1}{\milli\second} after inject.)}
    \end{subfigure}
    \caption{Longitudinal phase space $(\phi, E)$ reconstructed via tomography at different stages in the PSB (measurements).}
    \label{fig: how to hollow bunches}
\end{figure}

Eventually, the long filament can be smoothed to a ring-like phase space
distribution by high frequency phase modulation at harmonic $h=9$ with the C16
cavities. Figure \ref{fig: how to hollow bunches} shows tomographic
reconstructions \cite{tomo} of longitudinal phase space at important cycle times.
The horizontal axis is reverted compared to Fig.\ \ref{fig: simulations},
since $\phi=-z/R$ with $R$ the machine radius.

\section{Space Charge Mitigation in PS}
To assess the impact of direct space charge during the \SI{1.2}{\second}
PS injection plateau at $E_\text{kin}=\SI{1.4}{\giga\electronvolt}$ and $h=7$,
we compare single bunch beams of the usual LHC parabolic type with the modified
hollow type by measuring transverse emittance blow-up and beam loss.
For each shot, tomography and wire scans yield the $z,\delta, x, y$
distributions \SI{15}{\milli\second} after injection and again
\SI{20}{\milli\second} before the second batch injection time.
Table \ref{tab: ps parameters} lists the experiment parameters.
\begin{table}[htbp] \centering
    \caption{Relevant Experiment Beam Parameters for PS}
    \label{tab: ps parameters}
    \renewcommand{\arraystretch}{1.2}
    \begin{tabular}{ccc}
        \hline\hline
        \hbox{\hspace{-0.9cm}parameter\hspace{-1cm}} & hollow value & parabolic value \\ \hline
        $N$ & $(1.66 \mathrel{\pm} 0.05) \times 10^{12}$ & $(1.84 \mathrel{\pm} 0.03) \times 10^{12}$ \\
        $\epsilon_{z,100\%}$ & $1.43 \pm 0.15\,\si{\electronvolt\second}$ & $1.47 \pm 0.11\,\si{\electronvolt\second}$ \\
        $\epsilon_{z,\text{rms}}$ & $0.32 \mathrel{\pm} 0.02 \,\si{\electronvolt\second}$
            & $0.3 \mathrel{\pm} 0.01 \,\si{\electronvolt\second}$ \\ \hline
        $Q_x, Q_y$ & \multicolumn{2}{c}{$(6.23, 6.22)$} \\ \hline \hline
    \end{tabular}
\end{table}

\begin{figure}[htb]
    \includegraphics[width=\linewidth]{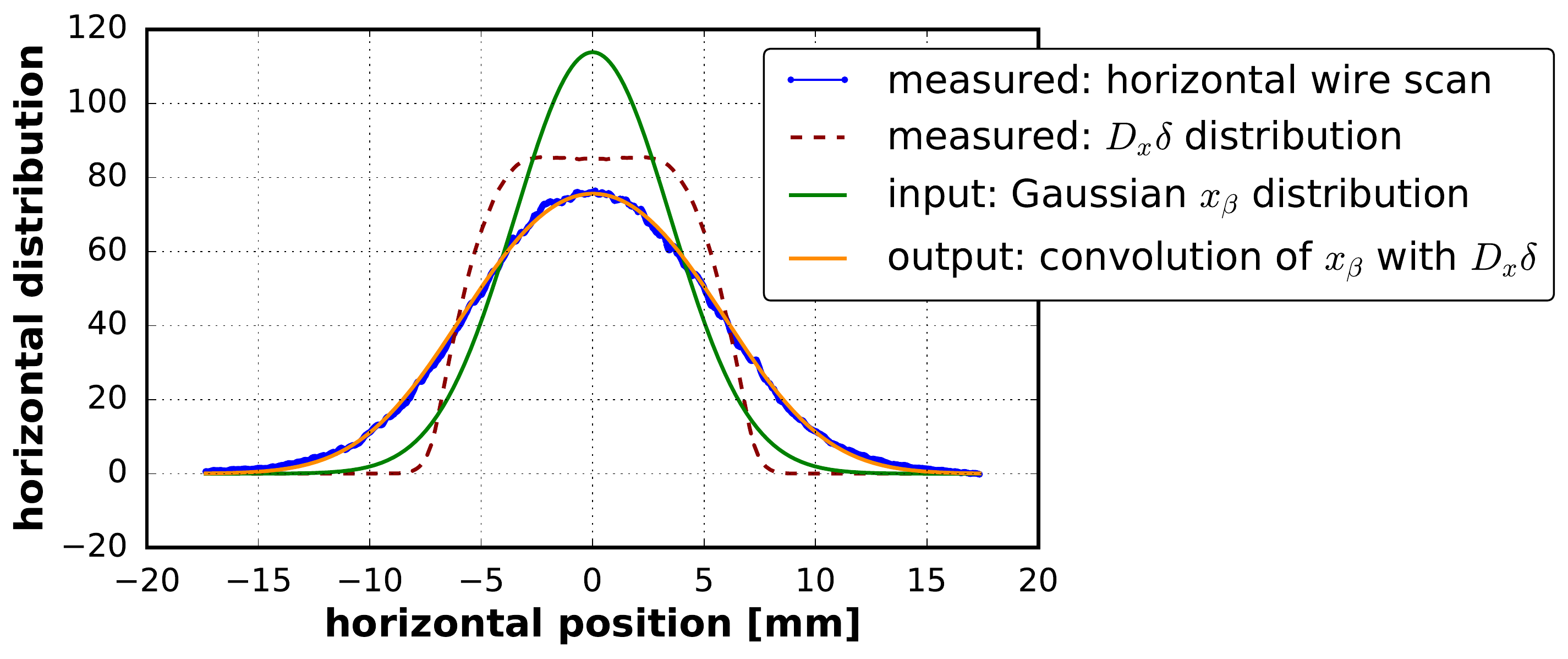}
    \caption{Wire scan comprising betatron and dispersive part.}
    \label{fig: convolution}
\end{figure}

Determining the horizontal emittance $\epsilon_x$ requires special attention:
since the momentum $\delta$ is by construction not Gaussian distributed for
hollow bunches, Eq.\ \eqref{eq: gauss beam size} is not valid.
The horizontal particle position is a sum of
two independent random variables, $x=x_\beta + x_{D_x \delta}$.
The dispersive profile is given by the measured $\delta$ distribution and
$D_x=\SI{2.3}{\meter}$ at the wire scanner location.
Convolving with a Gaussian distributed betatron profile hence
yields an estimate of the horizontal profile. The horizontal emittance
$\epsilon_x$ can then be found
by a least squares algorithm comparing the resulting convolution with the
actual measured profile, cf.\ Fig.\ \ref{fig: convolution}.
This procedure is applied to both beams. Results differ by $24.8\%$ to
$34.8\%$ from Eq.\ \eqref{eq: gauss beam size} in both cases.

\begin{figure}[htb] \centering
    \newcommand{\sfigheight}{5cm}
    \begin{subfigure}[t]{\linewidth} \centering
        \includegraphics[height=\sfigheight]{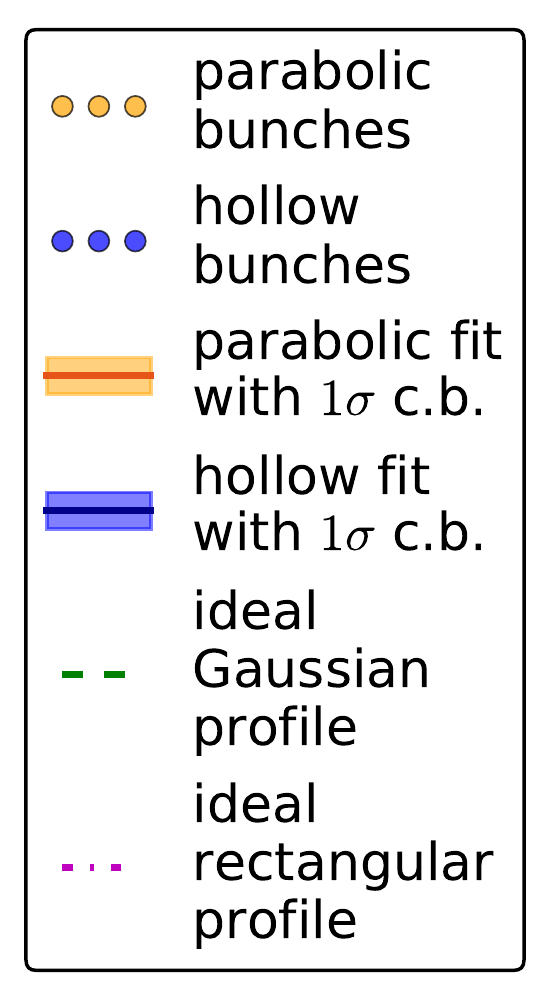} %
        \includegraphics[height=\sfigheight]{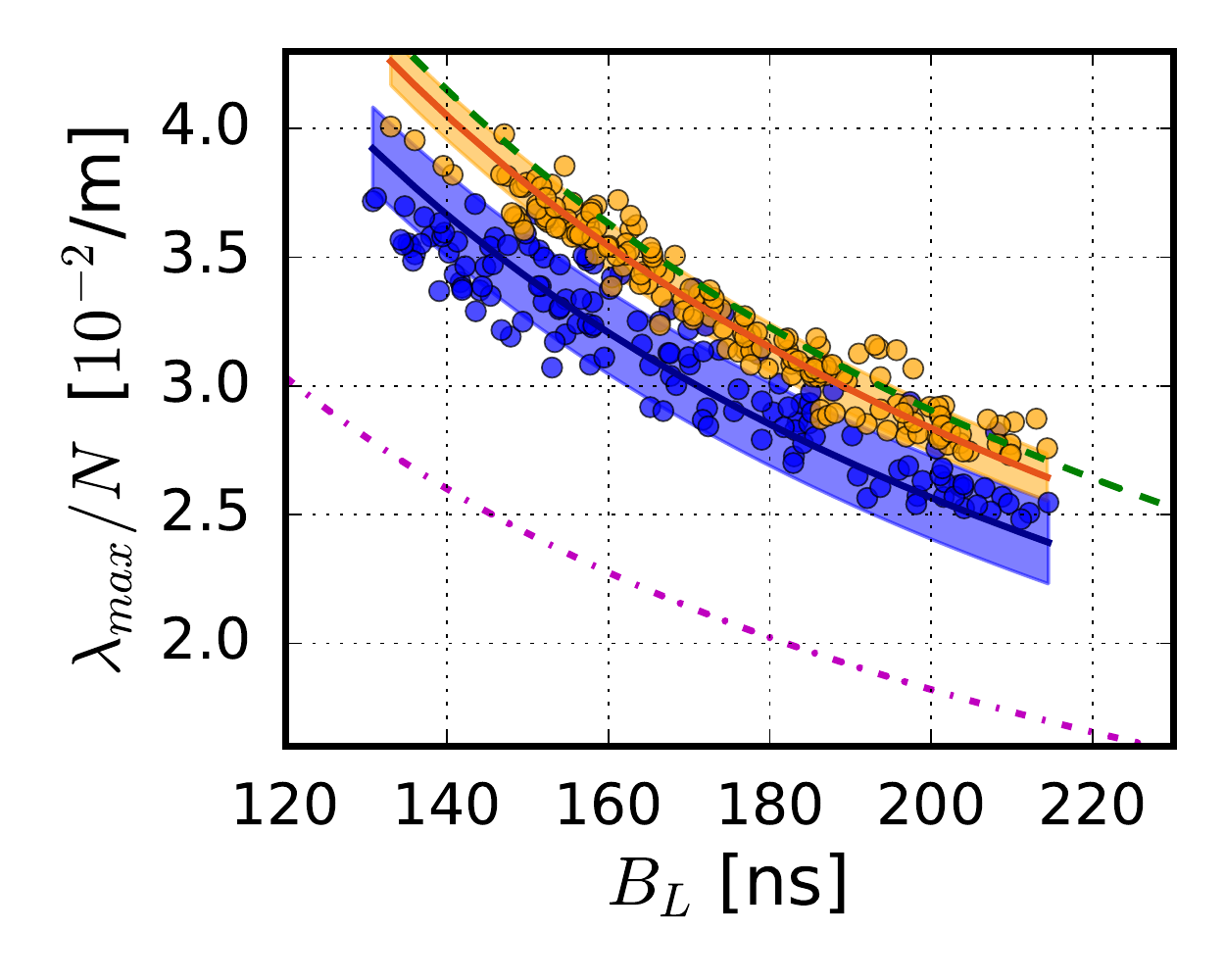}
        \begin{minipage}{0.85\linewidth} \centering
            \caption{Intensity normalised peak line charge density vs.\ total bunch length.}
            \label{fig: lambdamax}
        \end{minipage}
    \end{subfigure}
    \\
    \begin{subfigure}[t]{0.48\linewidth} \centering
        \includegraphics[height=\sfigheight]{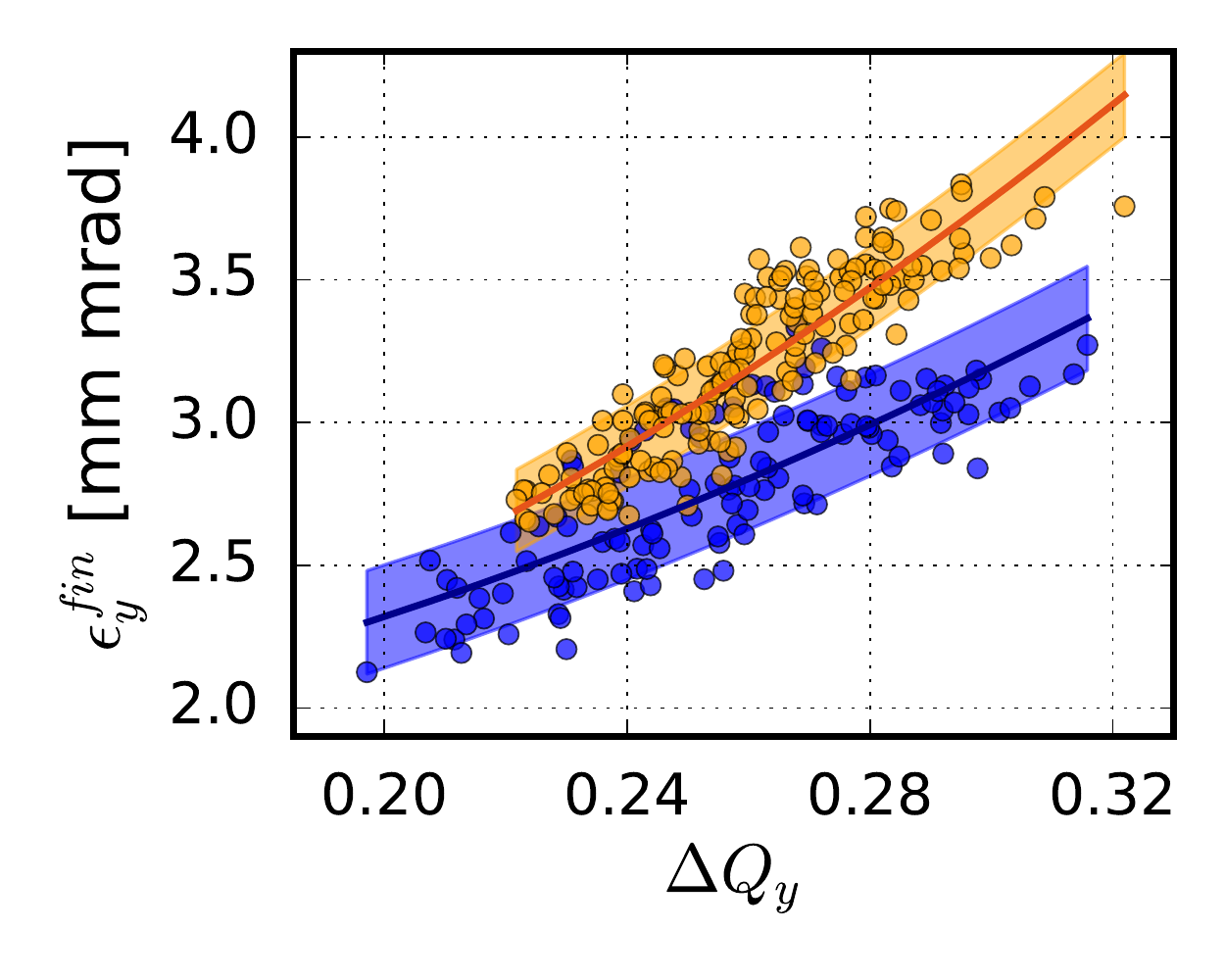}
        \begin{minipage}{\linewidth} \centering
            \caption{Vertical emittances (end of inject.\ plateau) vs.\ space
            charge tune shift.}
            \label{fig: finalcoreemit}
        \end{minipage}
    \end{subfigure}
    \hfill
    \begin{subfigure}[t]{0.48\linewidth} \centering
        \includegraphics[height=\sfigheight]{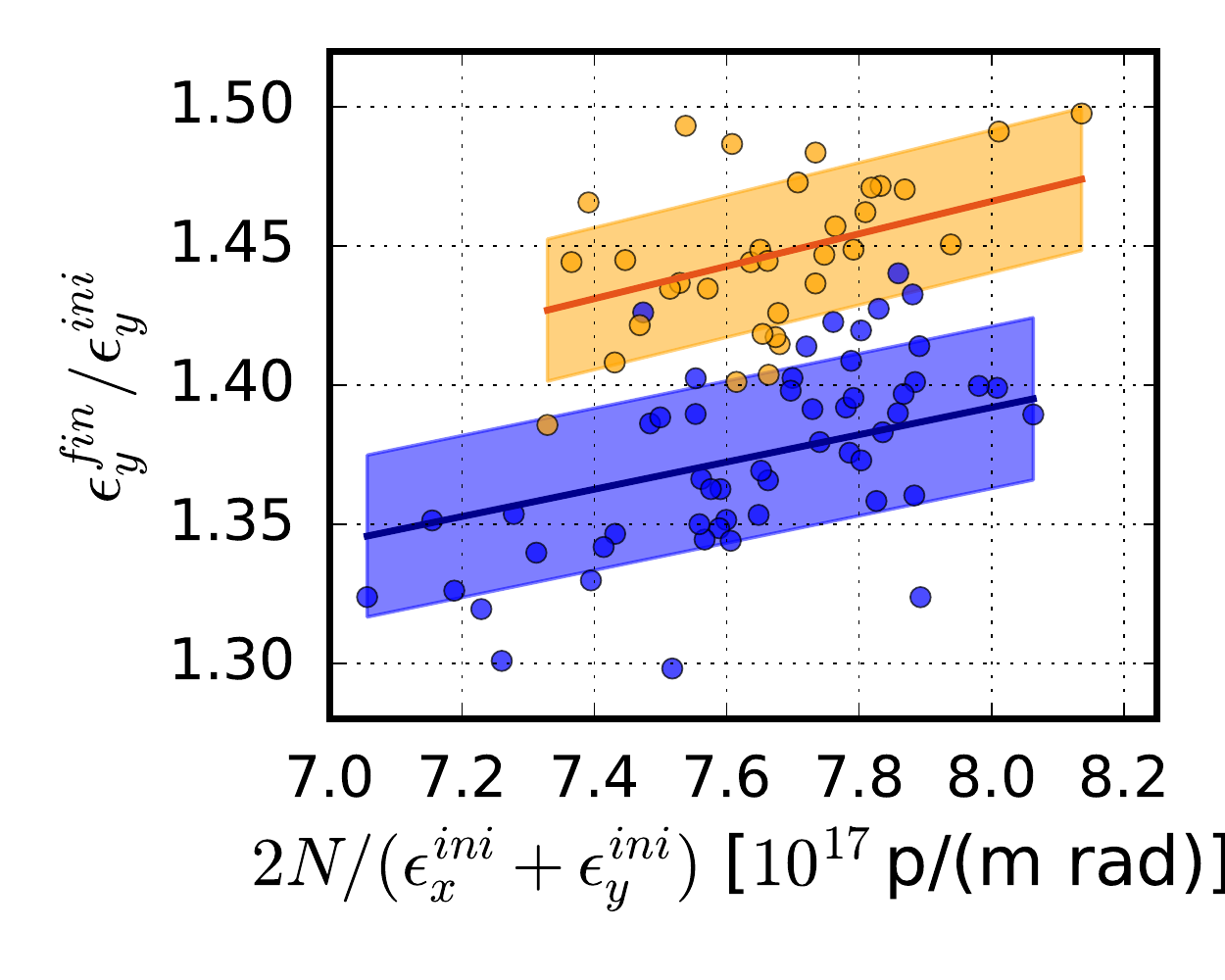}
        \begin{minipage}{0.85\linewidth} \centering
            \caption{Vertical emittance blow-up vs.\
            brightness (at $V_\text{rf}=\SI{80}{\kilo\volt}$).}
            \label{fig: blowup vs brightness}
        \end{minipage}
    \end{subfigure}
    \caption{Comparison of hollow and parabolic bunches. Fits include $1\sigma$ confidence bands.}
\end{figure}

Over many shots, we vary the bunch length for both beam types by
adiabatically ramping the total RF voltage during the
initial \SI{15}{\milli\second} to values between the initial \SI{25}{\kilo\volt}
and \SI{80}{\kilo\volt}. Due to varying shot-to-shot efficiency of the C16
blow-up, we achieve total bunch lengths over a range of
$B_L=130\, ..\,\SI{220}{\nano\second}$. Figure \ref{fig: lambdamax}
depicts consistently depressed peak line densities by a factor $0.9$
for the flattened profiles compared to the parabolic ones.
A theoretically ideal rectangular profile of $4\sigma_z$ length would yield
a $\sqrt{2\pi}/4\approx 0.63$ depression factor compared to a perfect Gaussian.
Both extrema are plotted in Fig.\ \ref{fig: lambdamax}
for comparison.

We want to compare the impact of space charge for both beam types for fixed
$B_L$, $N$ and $\epsilon_u$. To unify this set in one quantity, we choose to evaluate
$\Delta Q_u^\text{max}$ assuming a 6D Gaussian distributed beam in Eq.\
\eqref{eq: gauss tune spread}. Hence we apply \eqref{eq: gauss beam size} as
well as using the Gaussian peak line density
$\lambda_\text{max} = N / (\sqrt{2\pi} \sigma_z)$ where we set $\sigma_z=B_L/4$.
Figure \ref{fig: finalcoreemit} shows how hollow bunches provide statistically
significantly lower vertical emittances for the same unified reference tune
shift $\Delta Q_u^\text{max}$. The real tune shift of the hollow bunches is a
factor $0.88$ lower due to their reduced $\lambda_\text{max}$ and the
larger $\sigma_x$. In contrast, the parabolic bunches are rather well represented by
the Gaussian approach (factor $0.97$ lower real tune shift).

Finally, keeping the maximum RF voltage \SI{80}{\kilo\volt}, we scan the intensity by
varying the injected turns in the PSB. Figure \ref{fig: blowup vs brightness}
exhibits the emittance blow-up $\epsilon_y^\text{fin} / \epsilon_y^\text{ini}$
versus the brightness, which is again lower for the hollow bunches.

\clearpage

\section{Conclusion}
We have set up a reliable process to create hollow bunches with minimal
changes to the operational PSB cycle. Due to the lower peak line density,
the longitudinally hollow bunches are shown to be less affected by space charge
compared to the LHC-type parabolic bunches during the PS injection plateau.

\section{Acknowledgement}
The PSB feedback systems have been set up to create and transfer hollow bunches
owing to the invaluable support by Maria-Elena Angoletta and Michael Jaussi.


\end{document}